\documentclass[aps,prd,twocolumn,superscriptaddress]{revtex4}
\usepackage{epsfig,epsf}
\usepackage{amsmath}
\usepackage{amsthm}
\usepackage{amsfonts}
\usepackage{amssymb}
\usepackage{dsfont}
\usepackage{multirow}
\usepackage{appendix}
\usepackage{slashed}
\usepackage[active]{srcltx}
\usepackage{psfrag}

\begin{document}

\title{On the nature of the newly discovered $\Omega_c^{0}$ states}
\date{\today}
\author{S.~S.~Agaev}
\affiliation{Institute for Physical Problems, Baku State University, Az--1148 Baku,
Azerbaijan}
\author{K.~Azizi}
\affiliation{Department of Physics, Do\v{g}u\c{s} University, Acibadem-Kadik\"{o}y, 34722
Istanbul, Turkey}
\author{H.~Sundu}
\affiliation{Department of Physics, Kocaeli University, 41380 Izmit, Turkey}

\begin{abstract}
The masses and residues of the radially excited  heavy $\Omega_c^{0}$ and $%
\Omega_b^{-}$ baryons with spin-parity $J^{P}=\frac{1}{2}^{+}$ and $J^{P}=%
\frac{3}{2}^{+}$ are calculated by means of QCD two-point sum rule method
using the general form of their interpolating currents. In calculations the
quark, gluon and mixed vacuum condensates up to ten dimensions are taken
into account. In $\Omega_c^{0}$ channel a comparison is made with the narrow excited
states recently observed by the LHCb Collaboration \cite{LHCb}.
\end{abstract}

\maketitle

Recently, the LHCb Collaboration reported on discovery of new five narrow
states $\Omega _{c}^{0}$ in the $\Xi _{c}^{+}K^{-}$ invariant mass
distribution based on the $pp$ collision data at center-of-mass energies $7$%
, $8$ and $13$ $\mathrm{TeV}$ with an integrated $3.3b^{-1}$ luminosity \cite{LHCb}.
The masses of the $\Omega _{c}^{0}$ states  were measured and found to be
equal to (in $\mathrm{MeV}$) $M=3000;\,\,\,3050;\,\,\,3066;\,\,\,3090;\,\,\ 3119$. The LHCb determined also their widths through $\Omega
_{c}^{0}\rightarrow \Xi _{c}^{+}K^{-}$ decay channels, which did not
exceed a few $\mathrm{MeV}$.

Till now the experimental information on the spectrum of the charmed baryons
was limited by the $\Omega _{c}^{0}$ and $\Omega _{c}(2770)^{0}$ particles
with the masses \cite{Olive:2016xmw}
\begin{equation}
m=2695.2\pm 1.7\,\mathrm{\,MeV},\,\,m=2765.9\pm 2.0\,\,\mathrm{MeV}.
\label{eq:Data}
\end{equation}%
They are presumably the ground states $\Omega _{c}^{0}$ and $\Omega
_{c}^{0\star }$ with spin-parity $J^{P}=1/2^{+}$ and $J^{P}=3/2^{+}$,
respectively .

Theoretical investigations of the charmed (in general, heavy flavored)
baryons, on the contrary embrace variety of models and methods. The spectra
of the ground and excited states of the charmed baryons were analyzed in the
context of the QCD sum rule method \cite%
{Bagan:1991sc,Bagan:1992tp,Huang:2000tn,Wang:2002ts,Wang:2007sqa,Wang:2008hz,Wang:2009cr,Chen:2015kpa, Chen:2016phw}%
, different relativistic and non-relativistic quark models \cite%
{Capstick:1986bm,Ebert:2007nw,Ebert:2011kk,Garcilazo:2007eh,Valcarce:2008dr,Roberts:2007ni,Vijande:2012mk,Yoshida:2015tia,Shah:2016nxi}%
, the Heavy Quark Effective Theory (HQET) \cite{Chiladze:1997ev}, and in
lattice simulations \cite{Padmanath:2013bla}. The masses and magnetic
moments, radiative decays, various strong couplings and transitions of the
heavy flavored baryons were subject of rather intensive theoretical studies,
as well \cite%
{Aliev:2008sk,Aliev:2009jt,Aliev:2010nh,Aliev:2010ev,Aliev:2011kn,Aliev:2011ufa,Aliev:2011uf}%
. Some of these theoretical works were carried out using additional
assumptions on the structure of the charmed (bottom) baryons. For example,
in the relativistic quark model they were considered as the
heavy-quark-light-diquark bound states \cite{Ebert:2007nw,Ebert:2011kk}. In
other papers, QCD sum rule calculations aimed to evaluate spectroscopic
parameters of the charmed baryons were performed in the framework of HQET
\cite{Huang:2000tn,Wang:2002ts,Chen:2015kpa,Chen:2016phw}.

New experimental situation emerged due to the discovery of the LHCb
Collaboration, necessities a more detailed investigation of charmed (bottom)
baryons and their spectra. In the present Letter we will calculate
the masses and pole residues of the radially excited $2S$ charmed (bottom) $%
1/2^{+}$ and $3/2^{+}$ baryons in the framework of the  QCD full two-point sum rule
approach by employing the most general form of the interpolating currents
without any restricting suggestions about their internal organization. We are going
to follow a scheme applied to calculate the masses and residues
of the radially excited octet and decuplet baryons in Refs. \cite{Aliev:2016jnp,Aliev:2016adl}.
In these works, the authors get results, which are comparable with existing experimental data 
on the masses of the radially excited baryons, and  demonstrate that the QCD sum 
rule method can be successfully applied to investigate the radially excited baryons besides their
ground-states.

In order to derive the sum rules for the mass and residue of the radially
excited  $1/2^{+}$ and $3/2^{+}$ baryons we start from the two-point correlation
function
\begin{equation}
\Pi_{(\mu\nu)} (p)=i\int d^{4}xe^{ipx}\langle 0|\mathcal{T}\{J_{(\mu)}(x)\bar{J}_{(\nu)}(0)\}|0\rangle ,  \label{eq:CorrF1}
\end{equation}%
where $J(x)$ is the interpolating current for the  $\Omega _{c}^{0}$ ($%
\Omega _{b}^{-}$) states with $J^P=1/2^{+}$. It is given by the 
expression
\begin{eqnarray}
J &=&-\frac{1}{2}\epsilon ^{abc}\left\{ \left( s^{aT}CQ^{b}\right)
\gamma _{5}s^{c}+\beta \left( s^{aT}C\gamma _{5}Q^{b}\right) s^{c}\right.
\notag \\
&&\left. -\left[ \left( Q^{aT}Cs^{b}\right) \gamma _{5}s^{c}+\beta \left(
Q^{aT}C\gamma _{5}s^{b}\right) s^{c}\right] \right\} .  \label{eq:BayC1/2}
\end{eqnarray}%
In the case of the $3/2^{+}$ baryons the interpolating current has the form
\begin{eqnarray}
J_{\mu } &=&\frac{1}{\sqrt{3}}\epsilon ^{abc}\left\{ \left( s^{a}C\gamma
_{\mu }s^{b}\right) Q^{c}+\left( s^{a}C\gamma _{\mu }Q^{b}\right)
s^{c}\right.  \notag \\
&&\left. +\left( Q^{a}C\gamma _{\mu }s^{b}\right) s^{c}\right\}.
\label{eq:BayC3/2}
\end{eqnarray}
In expressions above $C$ is the charge
conjugation matrix, and $Q$ is the $c$ or $b$ quark. The current $J$ for the
$1/2^{+}$ baryons contains an arbitrary auxiliary parameter $\beta$, where
$\beta=-1$ corresponds to the choice of the Ioffe current.

The correlation function has to be calculated using both the physical and
quark-gluon degrees of freedom. To this end, we adopt the
``ground-state+first excited-state+continuum" scheme and calculate $\Pi ^{\mathrm{%
Phys}}(p)$ in terms of their parameters
\begin{equation}
\Pi ^{\mathrm{Phys}}(p)=\frac{\langle 0|J|\Omega _{Q}\rangle \langle \Omega
_{Q}|\bar{J}|0\rangle }{m^{2}-p^{2}}+\frac{\langle 0|J|\Omega
_{Q}^{\prime }\rangle \langle \Omega _{Q}^{\prime }|\bar{J}|0\rangle }{%
m^{\prime 2}-p^{2}}+\ldots ,  \label{eq:CF1/2}
\end{equation}%
where $\Omega _{Q}$ and $\Omega _{Q}^{\prime }$  are the ground and first radially
excited baryons with the masses $m$ and $m^{\prime }$, respectively. The
dots stand for contribution of the higher excited states and continuum.

Below we provide some details of calculations for the spin $1/2$ particles 
omitting, at the same time, corresponding expressions for the spin $3/2$ states. 
In order to proceed we introduce the matrix elements%
\begin{equation}
\langle 0|J|\Omega _{Q}^{(\prime )}(p,s)\rangle =\lambda ^{(\prime
)}u^{(\prime )}(p,s),\,  \label{eq:MElem}
\end{equation}%
where $\lambda $ and $\lambda ^{\prime }$ are the pole residues of the $%
\Omega _{Q}$ and $\Omega _{Q}^{\prime }$ states, respectively. Employing
Eqs.\ (\ref{eq:CF1/2}) and (\ref{eq:MElem}) and performing the summation
over the spins of the $1/2^{+}$ baryons
\begin{equation}
\sum\limits_{s}u^{(\prime )}(p,s)\overline{u}^{(\prime )}(p,s)=\slashed %
p+m^{(\prime )},
\end{equation}%
we get%
\begin{equation}
\Pi ^{\mathrm{Phys}}(p)=\frac{\lambda ^{2}(\slashed p+m)}{m^{2}-p^{2}}+\frac{%
\lambda ^{\prime 2}(\slashed p+m^{\prime })}{m^{\prime 2}-p^{2}}+\ldots
\label{eq:CorrF2}
\end{equation}%
The Borel transformation applied to Eq.\ (\ref{eq:CorrF2}) leads to the
result
\begin{equation}
\mathcal{B}\Pi ^{\mathrm{Phys}}(p)=\lambda ^{2}e^{-m^{2}/M^{2}}(\slashed %
p+m)+\lambda ^{\prime 2}e^{-m^{\prime 2}/M^{2}}(\slashed p+m^{\prime }).
\label{eq:Borl}
\end{equation}%
As is seen, there are two structures in Eq.\ (\ref{eq:Borl}), namely $\sim %
\slashed p$ and $\sim I$, both of which can be utilized to derive the
required sum rules%
\begin{eqnarray}
\lambda ^{2}e^{-m^{2}/M^{2}}+\lambda ^{\prime 2}e^{-m^{\prime 2}/M^{2}} &=&%
\mathcal{B}\Pi _{1}^{\mathrm{OPE}}(p),  \notag \\
\lambda ^{2}me^{-m^{2}/M^{2}}+\lambda ^{\prime 2}m^{\prime }e^{-m^{\prime
2}/M^{2}} &=&\mathcal{B}\Pi _{2}^{\mathrm{OPE}}(p),
\end{eqnarray}%
where $B\Pi _{1}^{\mathrm{OPE}}(p)$ and $B\Pi _{2}^{\mathrm{OPE}}(p)$ are
the Borel transformation of the corresponding structures in $\Pi (p)$ , but
calculated in terms of the quark-gluon degrees of freedom and labeled as $%
\Pi ^{\mathrm{OPE}}(p)$. Here we refrain from writing down rather lengthy 
expressions of the functions $B\Pi _{1}^{\mathrm{OPE}}(p)$ and $B\Pi _{2}^{\mathrm{OPE}}(p)$,
which will be published elsewhere.  

After rather simple manipulations we find the
desired sum rules for the parameters of the radially excited $\Omega
_{Q}^{\prime }$ state%
\begin{equation}
m^{\prime 2}=\frac{d/d(-1/M^{2})\left[ \mathcal{B}\Pi _{2}^{\mathrm{OPE}%
}(p)-m\mathcal{B}\Pi _{1}^{\mathrm{OPE}}(p)\right] }{\left[ \mathcal{B}\Pi
_{2}^{\mathrm{OPE}}(p)-m\mathcal{B}\Pi _{1}^{\mathrm{OPE}}(p)\right] },
\label{eq:SR1}
\end{equation}%
and%
\begin{equation}
\lambda ^{\prime 2}=\frac{e^{m^{\prime 2}/M^{2}}}{m^{\prime }-m}\left[
\mathcal{B}\Pi _{2}^{\mathrm{OPE}}(p)-m\mathcal{B}\Pi _{1}^{\mathrm{OPE}}(p)%
\right] .  \label{eq:SR2}
\end{equation}%
In the case of the $3/2^{+}$ baryons\ we use the matrix elements
\begin{equation}
\langle 0|J_{\mu }|\Omega _{Q}^{\ast (\prime )}(p,s)\rangle =\lambda
^{(\prime )}u_{\mu }^{(\prime )}(p,s),\,
\end{equation}%
where $u_{\mu }(p,s)$ are the Rarita-Schwinger spinors, and perform the
summation over the spins $s$ of the baryons by means of the formula
\begin{eqnarray}
&&\sum\limits_{s}u_{\mu }^{(\prime )}(p,s)\overline{u}_{\nu }^{(\prime
)}(p,s) =-(\slashed p+m^{(\prime )})\left[ g_{\mu \nu }-\frac{1}{3}\gamma
_{\mu }\gamma _{\nu }\right.  \notag \\
&&\left. -\frac{2}{3m^{(\prime )2}}p_{\mu }p_{\nu }+\frac{1}{3m^{(\prime )}}%
(p_{\mu }\gamma _{\nu }-p_{\nu }\gamma _{\mu })\right] .
\end{eqnarray}%
The interpolating current $J_{\mu }$ couples to the $1/2^{+}$ baryons, which
contribute to the sum rules, as well. Their contributions can be separated
and removed from the sum rules by special ordering of the Dirac matrices
(see, for example Ref.\ \cite{Aliev:2016jnp}). It is not difficult to
demonstrate, that the terms, which are formed exclusively due to
contribution of the $3/2^{+}$ baryons are proportional to the structures $\sim \slashed p
g_{\mu\nu}$ and $\sim g_{\mu\nu}$. Namely, these structures
and corresponding invariant amplitudes  are employed to get sum rules for the
masses and pole residues of the ground-state and radially excited $3/2^{+}$
charmed and bottom baryons.

The correlation function $\Pi ^{\mathrm{OPE}}(p)$ should be found using the
general expression Eq.\ (\ref{eq:CorrF1}) and heavy and light quarks'
propagators. In calculations we employ the $s$-quark and heavy quark
propagators given by the expressions%
\begin{eqnarray}
&&S_{s}^{ab}(x)=i\delta _{ab}\frac{\slashed x}{2\pi ^{2}x^{4}}-\delta _{ab}%
\frac{m_{s}}{4\pi ^{2}x^{2}}-\delta _{ab}\frac{\langle \overline{s}s\rangle
}{12}  \notag \\
&&+i\delta _{ab}\frac{\slashed xm_{s}\langle \overline{s}s\rangle }{48}%
-\delta _{ab}\frac{x^{2}}{192}\langle \overline{s}g_{s}\sigma Gs\rangle
+i\delta _{ab}\frac{x^{2}\slashed xm_{s}}{1152}  \notag \\
&&\times \langle \overline{s}g_{s}\sigma Gs\rangle -i\frac{%
g_{s}G_{ab}^{\alpha \beta }}{32\pi ^{2}x^{2}}\left[ \slashed x{\sigma
_{\alpha \beta }+\sigma _{\alpha \beta }}\slashed x\right]  \notag \\
&&-i\delta _{ab}\frac{x^{2}\slashed xg_{s}^{2}\langle \overline{s}s\rangle
^{2}}{7776}-\delta _{ab}\frac{x^{4}\langle \overline{s}s\rangle \langle
g_{s}^{2}G^{2}\rangle }{27648}+\ldots  \label{eq:qprop}
\end{eqnarray}%
and
\begin{eqnarray}
&&S_{Q}^{ab}(x)=i\int \frac{d^{4}k}{(2\pi )^{4}}e^{-ikx}\Bigg \{\frac{\delta
_{ab}\left( {\slashed k}+m_{Q}\right) }{k^{2}-m_{Q}^{2}}  \notag \\
&&-\frac{g_{s}G_{ab}^{\alpha \beta }}{4}\frac{\sigma _{\alpha \beta }\left( {%
\slashed k}+m_{Q}\right) +\left( {\slashed k}+m_{Q}\right) \sigma _{\alpha
\beta }}{(k^{2}-m_{Q}^{2})^{2}}  \notag \\
&&\left. +\frac{g_{s}^{2}G^{2}}{12}\delta _{ab}m_{Q}\frac{k^{2}+m_{Q}{%
\slashed k}}{(k^{2}-m_{Q}^{2})^{4}}+ \ldots \right\} .  \label{eq:Qprop}
\end{eqnarray}
\begin{table}[tbp]
\begin{tabular}{|c|c|}
\hline\hline
Parameters & Values \\ \hline\hline
$m_{b}$ & $4.18^{+0.04}_{-0.03}~\mathrm{GeV}$ \\
$m_{c}$ & $(1.27 \pm 0.03)~\mathrm{GeV}$ \\
$m_{s} $ & $96^{+8}_{-4}~\mathrm{MeV} $ \\
$\langle \bar{q}q \rangle $ & $-(0.24\pm 0.01)^3$ $\mathrm{GeV}^3$ \\
$\langle \bar{s}s \rangle $ & $0.8\ \langle \bar{q}q \rangle$ \\
$m_{0}^2 $ & $(0.8\pm0.1)$ $\mathrm{GeV}^2$ \\
$\langle \overline{s}g_{s}\sigma Gs\rangle$ & $m_{0}^2\langle \bar{s}s
\rangle $ \\
$\langle\frac{\alpha_sG^2}{\pi}\rangle $ & $(0.012\pm0.004)$ $~\mathrm{GeV}%
^4 $ \\ \hline\hline
\end{tabular}%
\caption{Parameters used in numerical computations.}
\label{tab:Param}
\end{table}
The correlation functions $\Pi _{1(2)}^{\mathrm{OPE}}(p)$ can be written
down in the form%
\begin{equation}
\Pi _{1(2)}^{\mathrm{OPE}}(p)=\int\limits_{(m_{Q}+2m_{s})^{2}}^{s_{0}}\frac{%
ds\rho _{1(2)}^{\mathrm{QCD}}(s)}{s-p^{2}},  \label{eq:SpecDen}
\end{equation}%
where $\rho _{1(2)}^{\mathrm{QCD}}(s)$ are the corresponding spectral
densities, and $s_0$ is the continuum threshold parameter. In Eq.\ (\ref%
{eq:SpecDen}) contribution of the higher exited states and continuum is
subtracted using the quark-hadron duality assumption.

As is seen, the sum rules for the excited states contain the mass of the
ground-state particles. We have evaluated the masses and pole residues of the
$1/2^{+}$ and $3/2^{+}$ ground-state baryons by employing the two-point sum
rule method within the ``ground-state + continuum" scheme. In calculations
the vacuum condensates up to ten dimensions are taken into account. The
masses of the ground-state baryons are used in Eqs.\ (\ref{eq:SR1}) and (\ref%
{eq:SR2}), and in the similar expressions for the $3/2^{+}$ baryons.

The sum rules depend on numerous parameters, which are collected in Table %
\ref{tab:Param}. It contains the masses of the bottom, charm and strange
quarks, as well as quark, gluon and mixed vacuum condensates. The sum rules
also require fixing of the working windows for the Borel parameter $M^2$ and
continuum threshold $s_0$, which are two auxiliary parameters of the
calculations. For $1/2^{+}$ particles we have additionally $\beta$ parameter coming
from the expression of the interpolating current. The choice of $M^2$, $s_0$, and $\beta$
is not totally arbitrary, but should satisfy the standard restrictions of the sum rule calculations.
Namely, the convergence of the operator product expansion, dominance of the pole
contribution, existence of the $(M^2,s_0)$ regions, where dependence of the
extracted quantities on $M^2$ and $s_0$ is minimal, have to be obeyed. The same is true for
$\beta$: we have to determine a working region for $\beta$ by demanding a weak dependence of our
results on its choice.

\begin{widetext}

\begin{table}[tbp]
\begin{tabular}{|c|c|c|c|c|}
\hline
$(n,J^{P})$ & $(1S,\frac{1}{2}^{+})$ & $(1S,\frac{3}{2}^{+})$ & $(2S,\frac{1}{2}^{+})$ & $(2S,\frac{3}{2}^{+})$\\ \hline
$M^2 ~(\mathrm GeV^2$) & $3.5-5.5$ & $3.5-5.5$ & $3.5-5.5$ & $3.5-5.5$ \\\hline
$s_0 ~(\mathrm GeV^2$) & $3.0^2-3.2^2$  & $3.1^2-3.3^2$ & $3.5^2-3.7^2$
& $3.6^2-3.8^2$ \\ \hline
$m_{\Omega_c} ~(\mathrm MeV)$ & $2685\pm 123$  & $2769 \pm 89$ & $3066 \pm 138$ &$3119 \pm 114$ \\ \hline
$\lambda_{\Omega_c} ~(\mathrm GeV^3)$ & $(6.2 \pm 1.8)\cdot 10^{-2}$   &$(7.1 \pm 1.0)\cdot 10^{-2}$& $(13.4 \pm 1.3)\cdot 10^{-2}$   & $(17.7 \pm 2.4)\cdot 10^{-2}$  \\ \hline
\end{tabular}%
\caption{The sum rule results for the masses and residues of the
$1S$ and $2S$  charmed baryons with the spin-parity $1/2^{+}$ and $3/2^{+}$.}
\label{tab:Results1A}
\end{table}
\begin{table}[tbp]
\begin{tabular}{|c|c|c|c|c|}
\hline
$(n,J^{P})$ & $(1S,\frac{1}{2}^{+})$ & $(1S,\frac{3}{2}^{+})$ & $(2S,\frac{1}{2}^{+})$ & $(2S,\frac{3}{2}^{+})$\\ \hline
$M^2 ~(\mathrm GeV^2$) & $6.5-9.5$ & $6.5-9.5$ & $6.5-9.5$ & $6.5-9.5$ \\\hline
$s_0 ~(\mathrm GeV^2$) & $6.3^2-6.5^2$  & $6.4^2-6.6^2$ & $6.8^2-7.0^2$
& $6.9^2-7.1^2$ \\ \hline
$m_{\Omega_b} ~(\mathrm MeV)$ & $6024 \pm 65$  & $6084\pm 84$ & $6325 \pm 92$ &$6412 \pm 108$ \\ \hline
$\lambda_{\Omega_b} ~(\mathrm GeV^3)$ &  $(12.1 \pm 1.2)\cdot 10^{-2}$ &$(9.3\pm 1.4)\cdot 10^{-2}$  & $(20.9 \pm 2.5)\cdot 10^{-2}$  &
$(23.6 \pm 3.4)\cdot 10^{-2}$  \\ \hline
\end{tabular}%
\caption{The $m_{\Omega_b}$ and $\lambda_{\Omega_b}$ of the ground-state and radially excited bottom baryons
with  $J^P=1/2^{+}$ and $J^P=3/2^{+}$.}
\label{tab:Results2A}
\end{table}

\end{widetext}

\begin{table}[tbp]
\begin{tabular}{|c|c|c|c|c|}
\hline\hline
$\Omega_c(1S,\frac{1}{2}^{+})$ & $\Omega_c^{\prime}(2S,\frac{1}{2}^{+})$ & $%
\Omega_c^{\star}(1S,\frac{3}{2}^{+})$ & $\Omega_c^{\star \prime}(2S,\frac{3}{%
2}^{+})$ &  \\
($\mathrm MeV$) & ($\mathrm MeV$) & ($\mathrm MeV$) & ($\mathrm MeV$) &  \\ \hline\hline
$2685 \pm 123$ & $3066 \pm 138$ & $2769 \pm 89$ & $3119 \pm 114$ & this work  \\ \hline
$2698$ & $3088$ & $2768$ & $3123$ & Ref.\ \cite{Ebert:2011kk} \\ \hline
$2699$ & $3159$ & $2767$ & $-$ & Ref.\ \cite{Valcarce:2008dr} \\ \hline
$2718$ & $3152$ & $2776$ & $-$ & Ref.\ \cite{Roberts:2007ni} \\ \hline
$2720\pm 180$ & $-$ & $2760 \pm 100$ & $-$ & Refs.\ \cite{Wang:2007sqa,Wang:2008hz} \\ \hline\hline
\end{tabular}%
\caption{The theoretical predictions for masses of the ground-state and
first radially excited $1/2^{+}$ and $3/2^{+}$ charmed baryons calculated in
the context of the different models and approaches.}
\label{tab:Lit1}
\end{table}
\begin{table}[tbp]
\begin{tabular}{|c|c|c|c|c|}
\hline\hline
$\Omega_b(1S,\frac{1}{2}^{+})$ & $\Omega_b^{\prime}(2S,\frac{1}{2}^{+})$ & $%
\Omega_b^{\star}(1S,\frac{3}{2}^{+})$ & $\Omega_b^{\star \prime}(2S,\frac{3}{%
2}^{+})$ &  \\
($\mathrm MeV$) & ($\mathrm MeV$) & ($\mathrm MeV$) & ($\mathrm MeV$) &  \\ \hline\hline
$6024 \pm 65$ & $6325 \pm 92$ & $6084 \pm 84$ & $6412 \pm 108$ & this work  \\ \hline
$6064$ & $6450$ & $6088$ & $6461$ & Ref.\ \cite{Ebert:2011kk} \\ \hline
$6056$ & $6479$ & $6079$ & $-$ & Ref.\ \cite{Valcarce:2008dr} \\ \hline
$6081$ & $6472$ & $6102$ & $-$ & Ref.\ \cite{Roberts:2007ni} \\ \hline
$6130\pm 120$ & $-$ & $6060 \pm 130$ & $-$ & Refs.\ \cite{Wang:2009cr,Wang:2008hz} \\ \hline\hline
\end{tabular}%
\caption{The various theoretical predictions for masses of the ground-state
and first radially excited $1/2^{+}$ and $3/2^{+}$ bottom baryons.}
\label{tab:Lit2}
\end{table}

Predictions obtained in this work for the masses and pole residues of the $1S$ and $2S$
charmed and bottom baryons with spin-parity $1/2^{+}$ and $3/2^{+}$, as well as the working
ranges of the parameters $M^{2}$ and $s_{0}$ are shown in Tables \ref%
{tab:Results1A} and \ref{tab:Results2A}, respectively. Results for the $(1S,1/2^{+})$ and
$(2S,1/2^{+})$   baryons are obtained by varying the parameter $\beta =\tan \theta $ in Eq.\ (\ref{eq:BayC1/2})  within the limits
\begin{equation}
-0.75\leq \cos \theta \leq -0.45,\  0.45\leq \cos \theta \leq 0.75,
\end{equation}%
to achieve the stable sum rules' predictions.

By comparison of  our results for  the ground state masses $m_{\Omega_c}$ and $m_{\Omega_c^{\star}}$
with the  experimental data given in Eq.\ (\ref{eq:Data}), we find a  nice agreement between them.
The same is correct for the bottom $\Omega_b^{-}$ baryon with spin-parity $J^P=1/2^{+}$ and  mass
$m=6071 \pm 40\,\mathrm{\,MeV} $ (see, Ref.\ \cite{Olive:2016xmw}).
Our predictions  allow us also to interpret the states $\Omega_c(3066)^{0}$  and $\Omega_c(3119)^{0}$,
recently discovered by the LHCb Collaboration, as the first radially excited $(2S, 1/2^{+})$
and $(2S, 3/2^{+})$ charmed baryons, respectively.

The spectroscopic parameters of the $\Omega_c$ baryons were calculated in the context of different
approaches, see Tables \ref{tab:Lit1} and \ref{tab:Lit2}. It is interesting
to note that our results for the charmed baryons are in accord with the predictions of the relativistic
diquark-quark model of Ref.\ \cite{Ebert:2011kk}. Other quark models lead almost to same predictions
for masses of the $\Omega_c^{0}$ and $\Omega_b^{-}$ baryons. Within systematic errors of the sum rule
computations we agree also with predictions of Refs.\ \cite{Wang:2007sqa,Wang:2009cr,Wang:2008hz} for the masses of ground-state
charmed and bottom baryons.

Investigations performed in the present Letter and results obtained in other works discussed above favor to assign
$\Omega_c(3066)^{0}$  and $\Omega_c(3119)^{0}$ states as the first radially excited $(2S, 1/2^{+})$
and $(2S, 3/2^{+})$ charmed baryons. Because LHCb did not fix the spin-parity of five $\Omega_c^{0}$ states,
the remaining three particles, namely $\Omega_c(3000)^{0}$, $\Omega_c(3050)^{0}$ and $\Omega_c(3090)^{0}$
maybe  $1P$-wave excitations of the charmed baryons. Our predictions on the ground-state and radially excited b-baryons with $J^{P}=1/2^{+}$ and $J^{P}=3/2^{+}$ may help experimentalists in the search for these states.

K.~A. thanks  Do\v{g}u\c{s} University for the partial financial support  through  the grant BAP
2015-16-D1-B04.

\end{document}